\documentclass[
reprint,
amsmath,amssymb,
aps,
]{revtex4-2}

\usepackage{graphicx}% Include figure files
\usepackage{dcolumn}% Align table columns on decimal point
\usepackage{bm}% bold math
\usepackage{booktabs}
\usepackage{amsmath,amssymb,amsfonts}
\usepackage{braket}
\usepackage{algorithmic}
\usepackage{textcomp}
\usepackage{xfrac}
\usepackage{verbatim}
\usepackage{hyperref}
\usepackage{xcolor}

\begin{document}

\preprint{APS/123-QED}

% \title{Spin mixing and polarization of single and double quantum transitions \\in photoexcited N-V defects in diamond}% Force line breaks with \\
% % \thanks{A footnote to the article title}%

% \title{Amplification and inversion of single and double quantum transitions in spin-mixed states under laser illumination}

\title{Amplification of single and double quantum transitions in spin-mixed states under laser illumination}

\title{Polarizing spin-mixed states in nitrogen-vacancy centers by photo-excitation}

\title{Single and double quantum transitions in spin-mixed states under photo-excitation}

\author{Anand Patel$^{1,2,\dag}$, Z. Chowdhry$^{1,3}$, Anil Prabhakar$^{2}$, A. Rathi$^{1,3}$, and V. P. Bhallamudi$^{1,2,3,\ddag}$}

\affiliation{$^{1}${Quantum Center of Excellence in Diamond and Emergent Materials, Indian Institute of Technology Madras, Chennai 600036, India}}
\affiliation{$^{2}${Department of Electrical Engineering, Indian Institute of Technology Madras, Chennai 600036, India}}
\affiliation{$^{3}${Department of Physics, Indian Institute of Technology Madras, Chennai 600036, India}\unskip}

\email[$^\dag$]{ee19d204@smail.iitm.ac.in}
\email[$^\ddag$]{praveen.bhallamudi@iitm.ac.in}

% \author{Anand Patel}
% \email{ee19d204@smail.iitm.ac.in} 
% \affiliation{Quantum Center of Excellence in Diamond and Emergent Materials, Indian Institute of Technology Madras, Chennai 600036, India}
% \affiliation{Department of Electrical Engineering, Indian Institute of Technology Madras, Chennai 600036, India}

% \author{Z. Chowdhry}
% \affiliation{Quantum Center of Excellence in Diamond and Emergent Materials, Indian Institute of Technology Madras, Chennai 600036, India}
% \affiliation{Department of Physics, Indian Institute of Technology Madras, Chennai 600036, India}

% \author{Anil Prabhakar}
% \affiliation{Department of Electrical Engineering, Indian Institute of Technology Madras, Chennai 600036, India}

% \author{A. Rathi}
% \affiliation{Quantum Center of Excellence in Diamond and Emergent Materials, Indian Institute of Technology Madras, Chennai 600036, India}
% \affiliation{Department of Physics, Indian Institute of Technology Madras, Chennai 600036, India}

% \author{V. P. Bhallamudi}
% \email{praveen.bhallamudi@iitm.ac.in} 
% \affiliation{Quantum Center of Excellence in Diamond and Emergent Materials, Indian Institute of Technology Madras, Chennai 600036, India}
% \affiliation{Department of Physics, Indian Institute of Technology Madras, Chennai 600036, India}
% \affiliation{Department of Electrical Engineering, Indian Institute of Technology Madras, Chennai 600036, India}

\date{\today}% It is always \today, today,

\begin{abstract}

Electronic spins associated with the Nitrogen-Vacancy (NV) center in diamond offer an opportunity to study spin-related phenomena with extremely high sensitivity owing to their high degree of optical polarization. Here, we study both single- and double-quantum transitions (SQT and DQT) in NV centers between spin-mixed states, which arise from magnetic fields that are non-collinear to the NV axis. We demonstrate the amplification of the ESR signal from  both these types of transition under laser illumination. We obtain hyperfine-resolved X-band ESR signal as a function of both excitation laser power and misalignment of static magnetic field with the NV axis. This combined with our analysis using a seven-level model that incorporates thermal polarization and double quantum relaxation allows us to comprehensively analyze the polarization of NV spins under off-axis fields. Such detailed understanding of spin-mixed states in NV centers under photo-excitation can help greatly in realizing NV-diamond platform's potential in sensing correlated magnets and biological samples, as well as other emerging applications, such as masing and nuclear hyperpolarization.  

% DOI:    PACS number(s): 

\end{abstract}
%Use showkeys class option if keyword %display desired
\maketitle

%\tableofcontents

\section{Introduction}

The negatively charged NV center in diamond \cite{schirhagl2014nitrogen} is a magneto-optically active defect that possesses an electronic spin-triplet (S = 1) with an exceptionally long spin lifetime \cite{balasubramanian2009ultralong}. A combination of remarkable properties make NV centers extremely promising for a wide range of applications such as magnetic sensing and imaging \cite{degen2008scanning,taylor2008high,bhallamudi2015nanoscale}, quantum communication \cite{dutt2007quantum,neumann2010quantum}, lasing \cite{savvin2021nv}, masing \cite{breeze2018continuous} and nuclear hyperpolarization \cite{jacques2009dynamic,busaite2020dynamic,duarte2021effect}. 
 
A key feature enabling these is the optical polarization of the NV spins, which can be orders of magnitude greater than thermal polarization even at room temperature. While a majority of work done with NV centers has used a static field collinear to NV-axis to maximize the signal, to truly develop the NV spin as a universal magnetic sensor, we should be able to apply off-axis field, \textit{i.e.}, fields not collinear with the NV axis. Expanding the scope of NV research with off-axis fields opens up new possibilities for sensing magnetic fields from various directions relative to the NV axis, facilitating three-dimensional mapping of magnetic field distributions. This is particularly significant in the study of correlated electron systems \cite{wolfe2014off, mccullian2020broadband}. Similarly, biosensing with nanodiamonds \cite{bhallamudi2015nanoscale,balasubramanian2014nitrogen,miller2020spin,teeling2016electron} may have off-axis fields given the challenges in controlling the orientation of nanodiamonds in biological samples. Understanding the dependence of polarization on field orientation may also be useful for NV-based amplifier or maser \cite{breeze2018continuous,sherman2021performance}, aiding in tunability and performance comprehension.

\begin{figure*}[htbp]
\centerline{\includegraphics[width=0.91\textwidth]{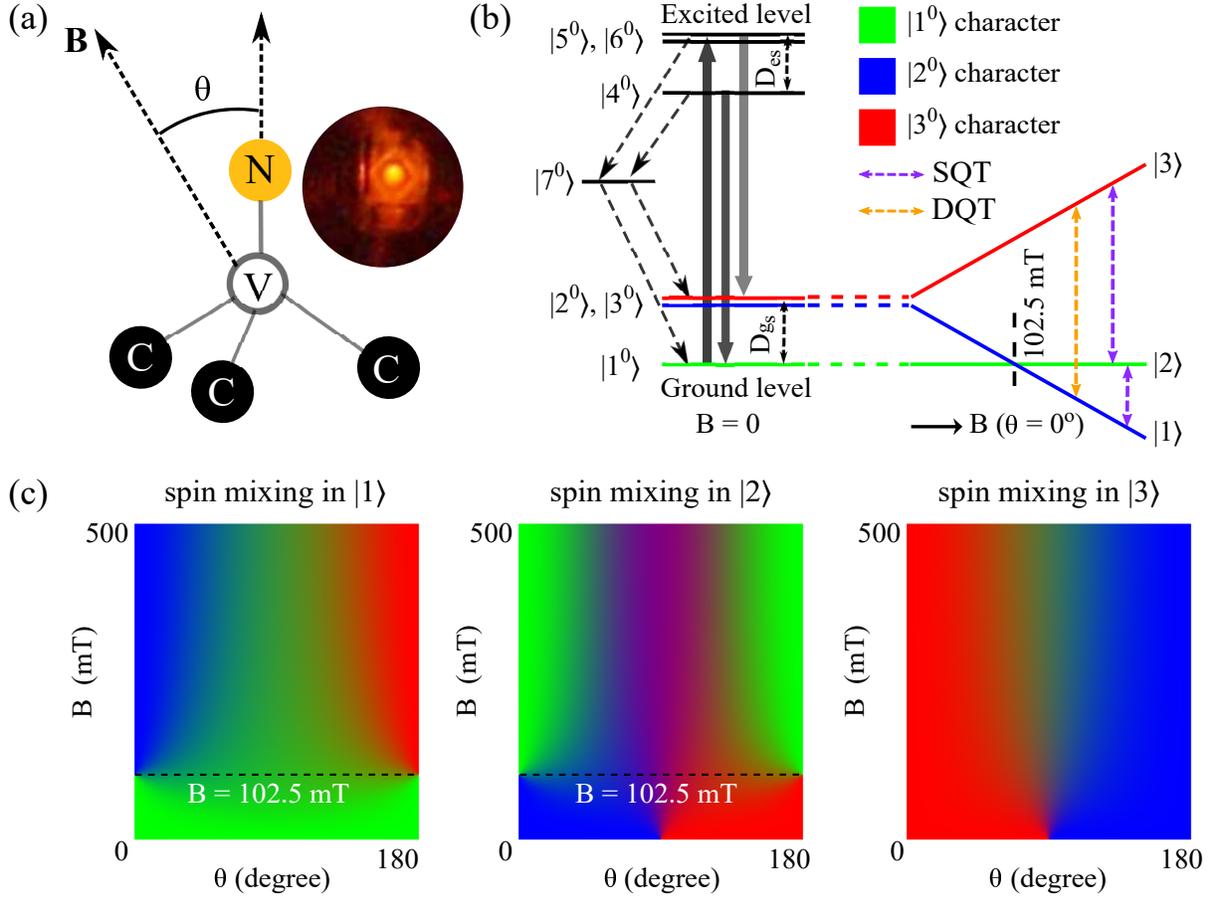}}  
    \caption{\textbf{Mixing of Nitrogen Vacancy (NV) spin sub-levels in seven-level (kinetic) model used in this study.} (a) Atomic configuration of the NV center with a substitutional Nitrogen (N) atom in adjacent to a vacancy (V) defect in the diamond lattice. The line joining the two, called NV-axis, is used as reference for degree of NV defect misalignment (polar angle $\theta$) with the applied magnetic field \textbf{B}. Inset shows the real-time (red) fluorescence image of NV-containing diamond single crystal under (green) laser illumination. (b) Schematic seven-level electronic structure of the NV center, showing optical excitation, spin-conserving radiative decay (solid lines) and nonradiative, spin-selective intersystem crossing (dashed lines) via the metastable singlet state $\ket{7}$. The labels with a superscript $^0$ represent the seven zero-field eigen states for each spin-pure sub-level ($m_s = 0, \pm1$). For the simplest case of aligned magnetic fields, \textbf{B}(B, $\theta$ = 0$^o$), a Zeeman splitting of the ground state (degenerate) spin sub-levels is also shown, with possible SQT (violet) and DQT (orange dashed lines) between them under (resonant) microwave excitation. DQT is only weakly allowed in $\theta$ = 0$^o$ case under local strain effects. (c) The calculated spin-mixing of (ground-state) spin sub-levels as a function of \textbf{B}(B, $\theta$). Green, blue, and red represent the $\ket{1^0}, \ket{2^0}, \ket{3^0}$ character or $\lvert\alpha_{i1}\rvert^2,\lvert\alpha_{i2}\rvert^2,\lvert\alpha_{i3}\rvert^2$ coefficients (see equation \eqref{alpha}) respectively where $\ket{i}$ is the given state.}
\label{Fig1}
\end{figure*}  

An off-axis (static) field also enables double ($\Delta m = 2$) quantum transitions (DQT) to be driven by ac magnetic/electric fields \cite{klimov2014electrically}, which is otherwise forbidden in aligned magnetic fields due to magnetic dipole selection rules. DQT offer increased sensitivity compared to conventional single quantum ($\Delta m = 1$) transitions (SQT). Furthermore, DQT exhibits more robustness against field misalignment (off-axis noise) \cite{moussa2014preparing} and strain effect \cite{felton2009hyperfine}. DQT, being magnetic-dipole-forbidden, can seectively probe the magnetic and electric noise \cite{myers2017double}. In the above context, studying the spin polarization of DQT under laser illumination while accounting for the extent of magnetic field misalignment provides an additional degree of freedom for NV-based diamond magnetometry.

% Ensemble samples are quite useful for wide-field magnetic imaging/sensing,  masing, and hyperpolazation. Understanding the polarization generated in NV centers by off-axis fields is important for these applications. Conventional inductively detected electron spin resonance experiments on ensembles can provide valuable information. Such measurement can focus on the effect of the off-axis fields on the polarization process, without the complication of changes in optical spin-readout due to off-axis fields, as in an optically detected magnetic resonance experiment.  

Optically induced spin polarization of NV has been studied in literature \cite{breeze2018continuous,sherman2021performance,drake2015influence} using conventional inductively detected electron spin resonance (ESR) experiments. Such measurement can specifically focus on the effect of the off-axis fields on the polarization process, without the complication of changes in optical spin-readout due to off-axis fields, as in an optically detected magnetic resonance experiment. Meanwhile, a comprehensive study analyzing the effect of both the knobs, off-axis field and laser intensity, is still lacking, especially for samples with low NV concentration ($<0.5$ ppm). Such samples, with extended spin lifetimes, are desirable for wide-field magnetic imaging applications.

In this work, we perform ESR spectroscopy on such a sample, studying the effects of varying laser intensity and magnetic field misalignment with the NV-axis. Our observations include both single and double quantum transitions. The experimental findings are interpreted  within the theoretical framework of NV spin mixing in its seven-level electronic structure. 

\vspace{-1 em}
\section{Spin mixing in NV centers}

NV center is an atom-like defect in diamond lattice (see figure \ref{Fig1}a) with the property of getting spin polarized by optical pumping with green laser. This occurs because of its seven-level electronic structure (see figure \ref{Fig1}b) where both the ground and excited states are spin triplets, and there is an additional singlet state ($\ket{7}$). The singlet state enables a non-radiative, spin-selective inter-system crossing, which play the central role in NV spin dynamics under laser illumination \cite{goldman2015state,thiering2018theory,duarte2021effect}. When optically excited, the $\ket{\mathrm{m_s}=0}$ preferentially decay in a spin-conserving radiative process, whereas $\ket{\mathrm{m_s}= \pm 1}$ also selectively populates $\ket{\mathrm{m_s}=0}$ ground state via singlet state ($\ket{7}$). This leads to the polarization of NV spins into the $\ket{\mathrm{m_s}=0}$ sublevel, even at room temperature, which is significantly higher than the polarization achievable through thermal equilibrium.

The ground-state triplet, whose spin population/polarization will be measured in the experiment, is described by following Hamiltonian:

\vspace{-1.0em}
\begin{equation}
\mathcal{H}_\text{gs} = hD_\text{gs} (\hat{S_{z}}^2 + \frac{2}{3}) + g\mu_\text{B} \mathbf{B\cdot \hat{S}} + \mathbf{\hat{S}A\hat{I}}
\label{ham}
\end{equation} 

where, $h$ is the Plank's constant, $D_\text{gs}=2.87~\text{GHz}$ is the zero-field splitting, $g\approx 2$ is the Land$\acute{e}$ g-factor and $\mu_\text{B}$ is the Bohr magneton. $\mathbf{B}~\text{and}~\mathbf{\hat{S}}$ are the applied magnetic field and spin vectors, respectively. The excited state is also described by a similar Hamiltonian $(\mathcal{H}_\text{es})$ with $D_\text{es}=1.42~\text{GHz}$. The zero-field term accounts for the magnetic anisotropy in the system. This defines the spin quantization axis to be along the NV-axis (see figure \ref{Fig1}a) which lies along one of the four $\braket{111}$ crystallographic directions in diamond lattice.
The third term represents the hyperfine interaction between the NV electron and \textsuperscript{14}N nuclei spin.  $\mathbf{A}~\text{and}~\mathbf{\hat{I}}$ are the nuclear hyperfine tensor and nuclear spin vector, respectively. We have ignored other terms in the Hamiltonian that are not important for the current study. 

In absence of external magnetic field, NV will have spin-pure states with $\mathrm{m_s}=0~(\ket{1^0},\ket{4^0}),~-1~(\ket{2^0},\ket{5^0})$ and $+1~(\ket{3^0},\ket{6^0})$ characters (see figure \ref{Fig1}b). 
When the magnetic field is precisely applied along the NV axis, specifically for \textbf{B}(B, $\theta$ = 0$^o$), the spin purity remains intact, except at the level anti-crossings at B = 102.5 mT, where the spin character gets swapped within the levels crossing over. Any off-axis field, \textbf{B}(B, $\theta$ $\neq$ 0$^o$) will result in mixing of the spin sub-levels. This spin mixing can be modelled by expressing each of the eigen-states as a linear combination of all the zero-field (spin-pure) eigen-states (see figure \ref{Fig1}b), as follows \cite{tetienne2012magnetic}. 
 
\vspace{-1.0em}
\begin{equation}
\ket{i} = \sum_{j=1}^{7}\alpha_{ij}(\mathbf{B})\ket{j^0}\label{alpha}    
\end{equation}

where, $\alpha_{ij}(\mathbf{B})$ are the complex coefficients. These can be numerically computed for both the ground and excited states using $\mathcal{H}_\text{gs}$ and $\mathcal{H}_\text{es}$, respectively. 

Figure \ref{Fig1}c represents the evolution of spin-pure state $i \in {1,2,3}$ for $\theta$ = 0$^o$ to a spin-mixed state in off-axis magnetic fields, \textbf{B}(B, $\theta$ $\neq$ 0$^o$) as a function of both field magnitude (B) and its orientation ($\theta$). The color indicates the proportion of the zero-field states. The strong spin-mixing, especially at fields and angles explored in this work, is clearly seen. 

\section{Results}

\begin{figure}
    \centering
    \includegraphics[width=0.45\textwidth]{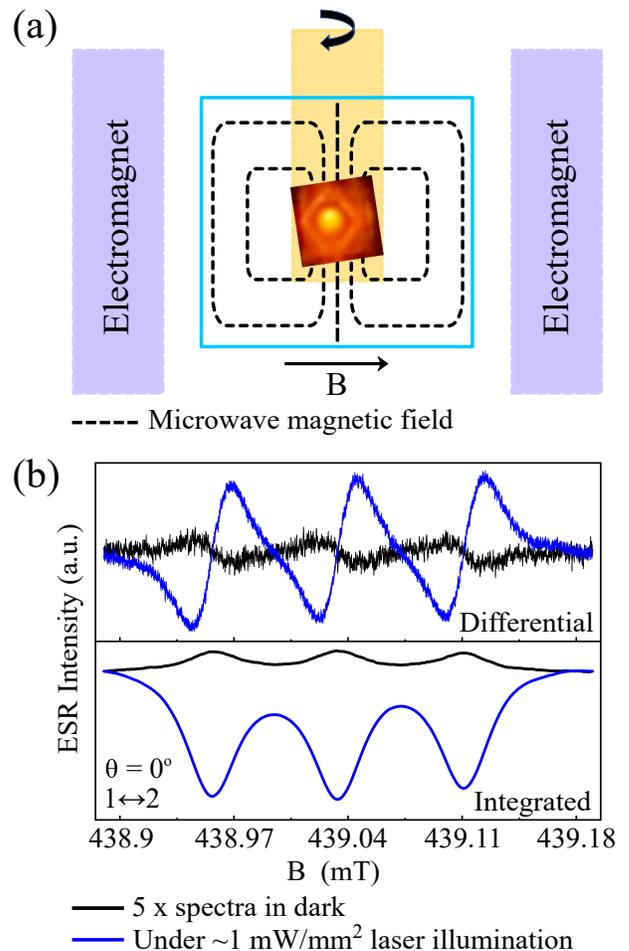}
    \caption{\textbf{Schematic for ESR spectroscopy of NV spin-mixed states under photo-excitation.} (a) The (100) diamond crystal was oriented at $45^\circ$ and glued to the quartz tube (yellow), which is inserted into a X-Band TE\textsubscript{011} cylindrical cavity (blue). It can then be rotated about its axis to achieve the desired degree of NV defect misalignment ($\theta$) with external \textbf{B} (see Section \ref{sec:methods} for details). The induced  spin-mixed states is studied by applying (resonant) microwaves (in dashed lines) in dark and under laser illumination via the optical access of the cavity. (b) Representative ESR spectra demonstrating the inversion and amplification of NV single-quantum ESR ($\ket{1}\leftrightarrow\ket{2}$) signal under photo-excitation.}
    \label{fig2:exp}
\end{figure}

\begin{figure*}
    \centering
    \includegraphics[width=0.92\textwidth]{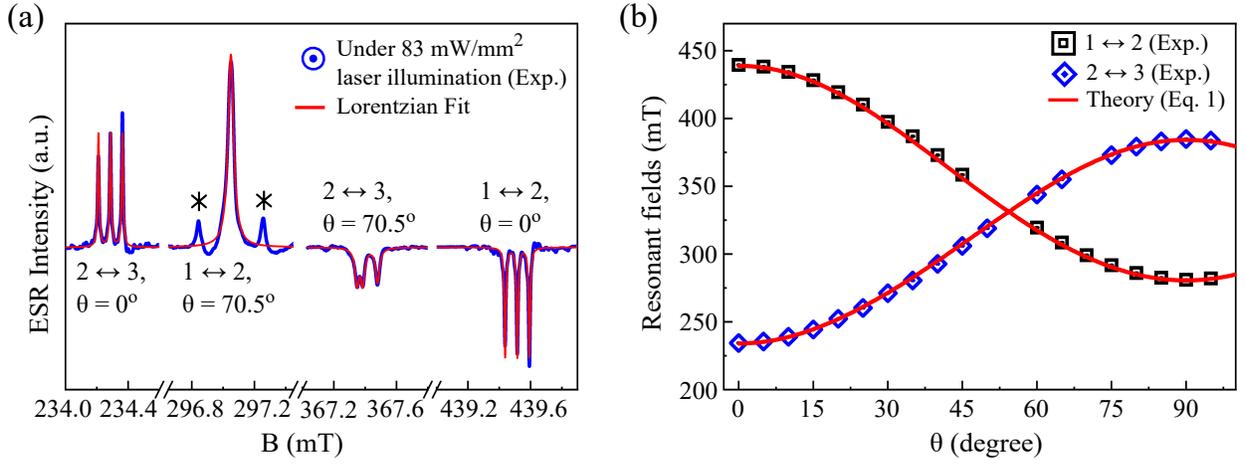}
    \caption{\textbf{Single-quantum ESR spectroscopy of NV spin-mixed states under laser illumination.} \textbf{(a)} (Integrated) single-quantum ESR spectra of NV centers (in blue symbols) measured under the maximum laser intensity of 83 mW/mm$^2$ at selected degrees ($\theta$ = 0$^o$ and $\simeq$ 70.5$^o$) of NVs orientation relative to \textbf{B} (see supplementary information for ESR spectra at all $\theta$ in figure S2). The ESR intensity is analyzed by fitting to multiple Lorentzian peaks (in red lines), associated with hyperfine splitting of the electronic transition. (b) The extracted resonant fields for the SQTs at all $\theta$ (in black and blue symbols) are compared to the theoretical values (in red lines) obtained by solving the ground state Hamiltonian (equation \eqref{ham}).
    }
    \label{fig3:ESR}
\end{figure*}

Figure 2a depicts the schematic of our experiment utilizing a commercial ESR spectrometer, featuring the capability of photoexciting NV spins during measurements. We measure the ESR spectra of a diamond sample containing NV centers, varying  their orientation ($\theta$) relative to the applied static magnetic field and laser illumination power. More details of the sample and experimental procedure are provided in Section \ref{sec:methods}: Methods. 

Figure \ref{fig2:exp}b displays the differential ESR spectra (top panel) acquired for the $\ket{1}\leftrightarrow\ket{2}$ SQT for $\theta$ = 0$^o$ with the laser illumination on and off. The spectra are integrated, and a background is removed (bottom panel) prior to further analysis (see supplementary information for baseline correction in figure S1). The three peaks correspond to the hyperfine levels of this transition resulting from the coupling between the NV electron spin and the \textsuperscript{14}N nuclei spin (I = 1). The linewidth analysis of the hyperfine peaks gives an upper bound on $T_2^*$ of $\simeq$ $5.6~\mathrm{\mu s}$, which is in close agreement with value provided by the sample manufacturer. 

The ESR spectrum obtained in the dark condition (with the laser turned off) represents the Boltzmann distribution of the  spin population in thermal equilibrium. Conversely, the second spectrum acquired under laser illumination at a low intensity of $\sim$ 1 mW/mm$^2$ displays a notable increase in the signal, which signifies the polarization of (ground state) spins driven by photo-excitation. This optically-induced polarization also results in inversion of the peaks \cite{breeze2018continuous}, reflecting a  population inversion (\textit{i.e.} higher spin population in the higher energy spin sublevel) as expected for the $\ket{1}\leftrightarrow\ket{2}$ transition above a field of 102.5 mT (see figure \ref{Fig1}b). ESR results, along with the deduced degree of spin polarization (with comparison to theory), for this and another SQT ($\ket{2}\leftrightarrow\ket{3}$) for $\theta$ = 0$^o$ and in off-axis magnetic fields (enabling  $\ket{1}\leftrightarrow\ket{3}$ DQT also) are described below.

\subsection{Single Quantum ESR Spectroscopy}

In figure \ref{fig3:ESR}a, the ESR spectra are shown for the experimental configuration with \textbf{B} pointing along one ($\theta$ = 0$^o$) of the four possible orientations of NV in diamond, under high-intensity laser illumination of 83 mW/mm$^2$. In such a case, the remaining NV centres in sample lies at $\theta$ = 70.5$^o$ relative to \textbf{B}. The two outermost peaks in the ESR spectra represents the $\ket{1} \leftrightarrow \ket{2}$ and  $\ket{2} \leftrightarrow \ket{3}$ SQTs for NVs at $\theta$ = 0$^o$, while the inner pair of peaks corresponds to the SQTs at $\theta$ = 70.5$^o$. In addition to $\ket{1} \leftrightarrow \ket{2}$ SQT at $\theta$ = 0$^o$, the $\ket{2} \leftrightarrow \ket{3}$ SQT at $\theta$ = 70.5$^o$ also show a population inversion (negative polarity), which is facilitated by spin-mixing in off-axis magnetic fields (see figure \ref{Fig1}b) as explored in subsection \ref{subsec:Polarization}. A key feature to notice is the unequal amplitudes of the hyperfine peaks at $\theta$ = 0$^o$, which indicates the capture of a notable polarization of the nuclear spins via the NV electron spin. Such a effect is minimal for hyperfine peaks under a much lower laser intensity of $\sim$ 1 mW/mm$^2$ (in figure \ref{fig2:exp}b). 

\begin{figure*}[htbp!]
    \centering
    \includegraphics[width=0.92\textwidth]{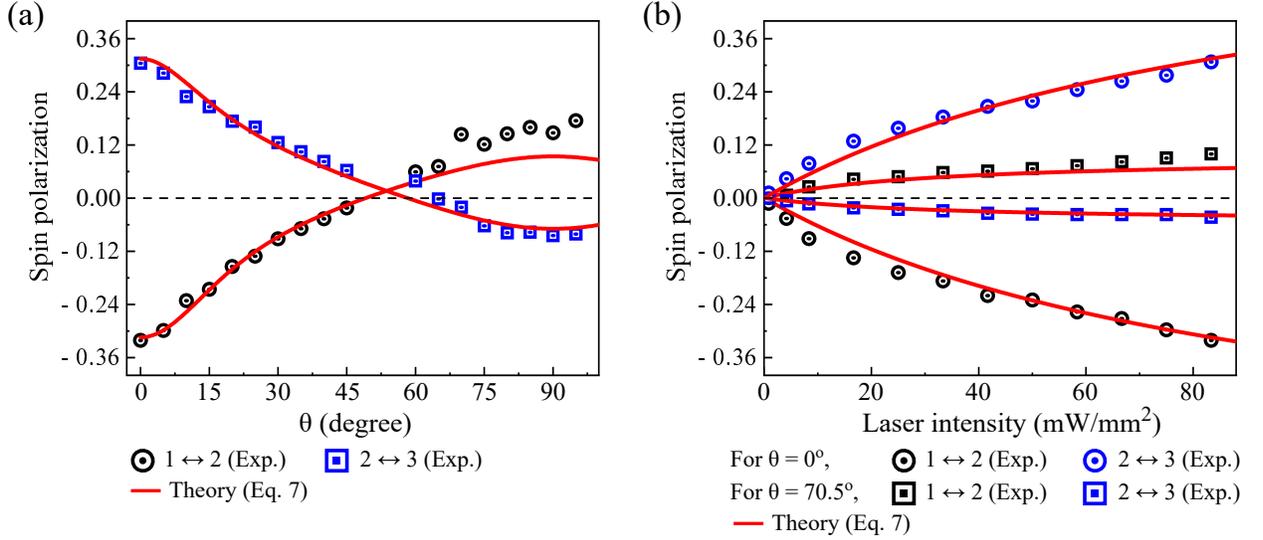}
    \caption{\textbf{Optically-induced spin polarization of single-quantum transitions in NV centers.} Spin polarization between the eigen-states of SQTs calculated from the ESR line intensities (in symbols) using equation \eqref{boltzman} as a function of (a) the degree of magnetic field misalignment, $\theta$ under the highest laser intensity of 83 mW/mm$^2$ and (b) the laser intensity for selected $\theta$ = 0$^o$ and 70.5$^o$. The experimental values are compared to the spin polarization values (in red lines) calculated from the seven level kinetic model using equation \eqref{ratmat} with parameters given in table \ref{tab:I}.
    }
    \label{fig4:SQT-SPz}
\end{figure*}

The picture becomes more complex in ESR signal measured for the case of $\theta$ = 70.5$^o$. The $\ket{1} \leftrightarrow \ket{2}$ and $\ket{2} \leftrightarrow \ket{3}$ SQTs show a marked difference in intensity, which can be attributed to a significant change in microwave coupling for two SQTs at that angle (see figure S3 in the supplementary information). Furthermore, the hyperfine peaks for $\ket{2} \leftrightarrow \ket{3}$ SQT tend to merge, while those for $\ket{1} \leftrightarrow \ket{2}$ merge into a single broader peak. This may arise from the imperfect experimental configuration (1$^\circ$ resolution of the setup) in orientation of NVs at the intended $\theta$ = 70.5$^o$ in three (possible) different directions, considering the strong variation of the resonant field with field orientation at this angle. The electronic transition peak (for $\ket{1} \leftrightarrow \ket{2}$) is accompanied by two additional peaks (marked by *), which may originate from the forbidden double-quantum nuclear transitions or hyper-polarization of the nuclear spins. 

We next acquired the ESR data by rotating the sample for the desired degree ($\theta$) of misalignment of \textbf{B}($\theta$) with the NVs in diamond, as described in section \ref{sec:methods}: Methods. The ESR data for both the SQTs at all $\theta$ (including $\theta$ = 0$^o$) is presented in figure S2 in the supplementary information. The ESR spectra are fitted to multiple Lorentzian peaks, with the constraint of equal magnitude for the hyperfine peaks (e.g., for $\theta$ = 0$^o$ and 70.5$^o$ in figure \ref{fig3:ESR}a). The (average) resonant fields obtained by fitting the data at all $\theta$ are plotted in figure \ref{fig3:ESR}(b), which agree well with the theoretically expected values by solving the Hamiltonian for NV spins in ground state (equation \eqref{ham}). This validates the Hamiltonian and subsequently spin-mixing calculations (see figure \ref{Fig1}c), which provide the basis for the analysis of degree of spin polarization in subsection \ref{subsec:Polarization} as follows.

\subsection{Spin Polarization}\label{subsec:Polarization}

We next focus on extracting the spin polarization, $S_z^{ij} = (n_i - n_j)/(\sum_{p=1}^7 n_p)$ between any two spin sublevels with eigen states $i,j$, with $n_i$ being the population of state $i$. Using the ESR data, the degree of spin polarization under laser illumination (with optical pumping strength  $\beta$) is calculated using the  relation as follows. 

\vspace{-1.0em}
\begin{equation}\label{boltzman}
% S_z^{ij}(\beta,\mathbf{B})= \frac{A_{ij}(\beta,\mathbf{B(\theta)})}{C_{ij}(\theta) A_{ij}(\beta =0, \mathbf{B}(\theta =0^o))} \cdot S_z^{ij}(\beta =0,\mathbf{B})
S_z^{ij}(\beta,\mathbf{B(\theta)})= \frac{A_{ij}(\beta,\mathbf{B(\theta)})}{C_{ij}(\theta)} \cdot \frac{S_z^{12}(0,\mathbf{B}(0^o))}{A_{12}(0, \mathbf{B}(0^o))} 
% = n_i-n_j
% \dfrac{\exp{\dfrac{-h\nu_0}{k_\text{B}T}}-1}{\exp{\dfrac{-h\nu_0}{k_\text{B}T}}+1}
\end{equation}

For the specific transition ($\ket{i} \leftrightarrow \ket{j}$), $A_{ij}$ represents the area under the peak (summed over all the hyperfine peaks) and $C_{ij}$ accounts the change in microwave coupling at that $\theta$ relative to that for $\ket{1}\leftrightarrow\ket{2}$ SQT at $\theta$ = 0$^\circ$, which can be found in Figure S3 in the supplementary information. $S_z^{12}(0,\mathbf{B}(0^\circ))$ refers to the thermal spin polarization, which is calculated by quantifying the Boltzmann's distribution of NV spins in two sublevels ($\ket{1}$, $\ket{2}$) using ESR data acquired for SQT between them at $\theta$ = 0$^\circ$ in dark condition ($\beta$ = 0), as presented in figure \ref{fig2:exp}b. The optically-induced spin polarization thus calculated, using equation \eqref{boltzman}, for two SQTs are plotted in figure \ref{fig4:SQT-SPz} with varying $\theta$ (at maximum laser intensity of 83 mW/mm$^2$) as well as the laser intensity (at specific $\theta = 0^\circ \text{ and } 70.5^\circ$). 

To make a quantitative understanding of optically-induced spin polarization in the framework of spin-mixing, we have also calculated it computationally using seven-level kinetic model (presented in figure \ref{Fig1}b) by extracting the steady state populations for spin-mized eigen states ($i,j$) from classical rate equation as follows. 

\vspace{-1.0em}
\begin{equation}
\frac{dn_i}{dt}=\sum_{j=1}^7(k_{ij}n_j-k_{ji}n_i)\label{rateq}
\end{equation}

\begin{figure*}
    \centering
    \includegraphics[width=0.98\textwidth]{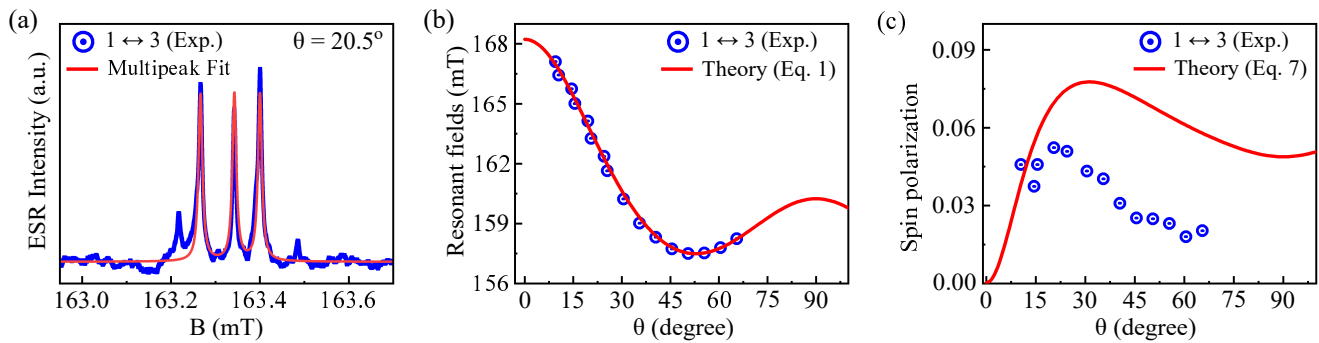}
    \caption{\textbf{Double-quantum ESR spectroscopy of NV centers in off-axis magnetic fields.} (a) (Integrated) ESR spectra under the laser illumination of $83~\mathrm{mW/mm^2}$, after baseline correction, of DQT between spin-mixed states at $\theta$ = 20$^o$ (in blue symbols), with fitting into multiple hyperfine peaks with a Lorentzian profile (in red line). (b) The extracted resonant fields at different $\theta$ are plotted against the theoretical values (in red line) derived from the ground state Hamiltonian (equation \eqref{ham}). (c) The experimentally calculated spin polarization (in symbols) using equation \eqref{boltzman} for DQT at different $\theta$ are compared to the values obtained from the kinetic calculations using equation \eqref{ratmat} with parameters given in table \ref{tab:I}.}
    \label{fig5:DQT-SPz}
\end{figure*}

Here, the optical transition rates, $k_{ij}(\mathbf{B})$ between spin-maxed states for the specific ($\ket{i} \leftrightarrow \ket{j}$) transition can be obtained using the following relation:

\vspace{-1.0em}
\begin{equation}
k_{ij}(\mathbf{B}) = \sum_{p=1}^{7}\sum_{q=1}^{7}\lvert\alpha_{ip}\rvert^2\lvert\alpha_{jq}\rvert^2k_{pq}^0\label{ktrans}
\end{equation}  

The coefficients, $\alpha_{ij}(\mathbf{B})$ are computed by solving the Hamiltonian (equation \eqref{ham}) at resonant fields for the respective  transition (see figure S4 in the supplementary information). Here $k_{pq}^0$ are the zero-field optical transition rates except $k_{21,12,31,13}^0 = 1/2T_1$ where $T_1$ is the longitudinal spin relaxation rate. While zero-field optical decay rates, including the intersystem crossing, are measured experimentally in literature (see Table I), the optical pumping rates can be calculated from the radiative decay rates ($k_r^0$) using the relation $k_{14,25,36}^0 = \beta k_\text{r}^0$. Here $\beta$ is a dimensionless parameter related to the optical pumping rate as follows.

\vspace{-1.0em}
\begin{equation}
\beta = \frac{\sigma}{4 \cdot k_\text{r}^0 \cdot h\nu} \times \text{laser intensity},\label{beta}
\end{equation}

 where $\sigma$ is the absorption cross-section of NV centres under 532 nm laser illumination, $\nu$ is the laser frequency. A factor of 4 accounts the optical excitation of NVs in only one of the four possible orientations in diamond.

In steady state, one of the classical rate equations become redundant and can be replaced with $\sum n_i = 1$. In turn, this normalizes the solution w.r.t. the total NV concentration (as in the definition of spin polarization). To account the thermal effect, equation \eqref{rateq} can be solved by expressing it in matrix form: 

\vspace{-1.0em}
\begin{equation}
\mathbf{A} n = B \label{ratmat}
\end{equation}

where $\mathbf{A}_{7 \times 7}$ is the matrix of transition rate coefficients, $k_{ij}$, while $n_{7 \times 1}$ is a column matrix of unknown populations. At a given temperature i.e., room temperature in current study, another column matrix,  $B_{7 \times 1}$ can be calculated by setting $\beta$ to 0 (dark) in transition rate matrix $A$ and replacing $n$ with the dark state populations given by the Boltzmann's distribution assuming that all the NV spins are in the ground state manifold at room temperature, thus $i~\text{or}~j \in \{1,2,3\}$.
% As desctibed by the folliowing equation:

\begin{table}[]
\begin{tabular}{@{}ll@{}}
\toprule
Parameter        & Value\\ 
\midrule
$k_{41,52,63}^0$ & 62.7 MHz \\
$k_{47,67}^0$    & 80 MHz \\
$k_{57}^0$       & 12.97 MHz \\
$k_{71,73}^0$    & 1.08 MHz \\
$k_{72}^0$       & 3.45 MHz \\
$T_1$            & 5.5 ms \\
% $T_1$ (SQ)       & 5.5 ms \\
% $T_1$ (DQ)       & 2.75 ms \\
$\sigma$       & $9.3\times 10^{-17}~\text{cm}^2$ \\ 
\bottomrule
\end{tabular}
\caption{Parameters used in seven-level kinetic model for calculation of spin polarization values using using equation \eqref{ratmat}. Zero-field transition rates are taken from \cite{robledo2011spin}.}
\label{tab:I}
\end{table}

Using the parameters provided in table \ref{tab:I}, the obtained spin polarization are plotted in figure \ref{fig4:SQT-SPz}. The 7-level energy model of NV centres successfully predicts the trend of angular and laser power dependence of spin polarization in experimental results. The quantitative agreement between simulated and experimental polarization is discussed in Section \ref{sec:discussion}.

\subsection{Double Quantum Transitions}
 
In the ESR data acquired for experimental configuration with NV centers in off-axis \textbf{B}(B, $\theta$ $\neq$ 0$^o$) under photo-excitation, we made a notable observation of emergence of the ESR signal corresponding to $\ket{1} \leftrightarrow \ket{3}$ DQT (e.g., for $\theta = 20^\circ$ in figure \ref{fig5:DQT-SPz}a). This finding is confirmed by comparing the resonant fields extracted from the ESR signal at all non-zero $\theta$ (through multipeak Lorentzian fits) with the values obtained by solving the Hamiltonian (equation \eqref{ham}) for this transition (see \ref{fig5:DQT-SPz}b). This is an intriguing finding because under perfect alignment ($\theta = 0^\circ$), this particular transition between $\mathrm{m_s} = -1~(\ket{1^0})$ and $+1~(\ket{3^0})$ (above 102.5 mT) is not allowed. It is only under the strong spin-mixing regime (see figure \ref{Fig1}c) where these transitions are allowed and are visible in the ESR spectra. Nevertheless, the observation of the DQT signal can provide valuable insights into the nature (angle) and strength of the misaligned magnetic fields relative to NV centers. This can help improving the performance of their various applications particularly quantum sensing, where precise information of the magnetic field is essential.   

We also calculated the optically-induced spin polarization, both, experimentally from the ESR data (using equation \eqref{boltzman}) and computationally by solving the seven-level rate equation into matrix formalism (using equation \eqref{ratmat}). The extracted values are shown in figure \ref{fig5:DQT-SPz}c. A maximum (experimental) spin polarization is obtained  within the $\theta$ range of 10$^\circ$-35$^\circ$. While qualitatively the trends of the experimental data (markers) and fits (solid lines) match well, we cannot obtain as good a quantitative fit for larger $\theta$ using the same parameters as the SQT case, as further discussed in Section \ref{sec:discussion}. 
% \textcolor{red}{This canbe likely due to...}

\section{Discussion and conclusions}\label{sec:discussion}

% The widely used seven-level model was extended by \cite{tetienne2012magnetic} to incorporate spin-mixing effects for off-axis magnetic fields. However, their focus was primarily on the low magnetic field regime and specific angles. Their simulations primarily aimed to study the radiative lifetimes as a function of spin-mixing without needing to consider the influence of thermal polarizations or spin polarization as a function of laser illumination.

The commonly used seven-level kinetic model of NV centers was extended by \cite{tetienne2012magnetic} to incorporate spin-mixing effects in off-axis magnetic fields. Authors primarily focus on the low magnetic field regime and specific angles. Further studies \cite{breeze2018continuous, sherman2021performance} have investigated NV spin polarization under photo-excitation with seven-level model analysis at high fields, but specifically for aligned fields. In other work, \cite{drake2015influence} examined the angular dependence of spin polarization in high off-axis magnetic fields. Their approach accounted for this by applying the Wigner rotation exclusively to the ground state Hamiltonian while considering equal population for $m_\text{s}=\pm1$ levels and assuming a fixed polarization for the aligned case based on experimental data. In contrast, seven-level kinetic model, further incorporating thermal polarization, employed by us also derive the aligned case spin polarization from first principles and considers the differences in zero-field splittings ($D_\text{gs}$ and $D_\text{es}$) for the ground and excited states, which gives different spin mixing conditions. This is evident from the observation of DQT in our ESR data measured in off-axis field, demonstrating the distinct populations of $m_\text{s}=\pm1$ sub-levels.

% Similarly, the work conducted by \cite{breeze2018continuous} and \cite{sherman2021performance} investigated ground state spin polarizations under room temperature photo-excitation but solely for aligned fields. On the other hand, \cite{drake2015influence} examined the angular dependence of spin polarization in off-axis magnetic fields, which they accounted by the Wigner rotation of only the ground state Hamiltonian. Additionally, they employed the initial polarization for aligned fields as a fitting parameter, rather than utilizing a kinetic model to derive population from first principles. Moreover, their approach did not account for the distinct spin-mixing conditions between the ground and excited state sub-levels, resulting from their differing zero-field splitting parameters and predicted equal population for $m_\text{s}=\pm1$ states. 

Moreover, none of the previous studies have investigated the amplification of double quantum transitions (DQT) under laser illumination for high off-axis field conditions. Our objective is to comprehensively explore the phase-space of these experimental parameters, enabling a more thorough understanding of the influence of laser illumination on ground state spins.

In this work, we conducted systematic ESR experiments on a diamond sample with a relatively lower NV concentration ($\approx$ 0.2 ppm) compared to previous such studies ($\geq$ 0.49 ppm in \cite{sherman2021performance} and $\geq$ 1.9 ppm in \cite{drake2015influence}). The lower concentration with long $T_2^\ast$ of $\simeq$ $5.6~\mathrm{\mu s}$ (vs 29 ns in \cite{sherman2021performance}) gives rise to fully resolved $^{14}$N hyperfine levels in our ESR data. This longer $T_2^\ast$ is advantageous for magnetic sensing applications. In contrast, the higher NV density samples used in \cite{sherman2021performance,drake2015influence} exhibited broader linewidths, which are more desirable for diamond maser applications, specifically to achieve the high bandwidth of operation, as highlighted by Sherman et al. \cite{sherman2021performance}. 

We compared our experimental results of optically-induced enhancement of spin polarization (vs thermal polarization) for $\theta = 0^\circ$ case with previous works. Our data demonstrated a high amplification of $\times$ 685 for SQTs at laser intensity of 83 mW/mm$^2$. This surpasses the maximum amplification of $\approx$ $\times$ 400 obtained Sherman et al. \cite{sherman2021performance} at similar illumination intensity for their lowest NV density (0.49 ppm) sample. The measured data in their study indicates a lowering of enhancement and $T_1$ as NV density increases in the sample. Subsequent kinetic model calculations, considering experimental $T_1$ values, clearly emphasize the noteworthy impact of $T_1$ on the spin polarization of the sample. Based on this analysis, the higher amplification in our low NV density sample compared to \cite{sherman2021performance} can be attributed to a relatively higher $T_1$ for our sample.

In other work, Drake et al. \cite{drake2015influence} reported a saturation of amplification, reaching $\approx$ $\times$ 350 at a much lower laser intensity of $\approx$ 30 mW/mm$^2$. While above-stated works, including our, are performed at room temperature, Degen and co-workers \cite{loretz2017optical} conducted ESR experiments at 200 K using a higher NV density sample (9 ppm). Their observations followed a similar trend as \cite{drake2015influence}, revealing a optical saturation effect at even lower laser intensity of $\approx$ 10 mW/mm$^2$ and a maximum enhancement of $\times$ 170. The team accounted this saturation effect in their kinetic model calculations by considering the NV (negatively charged $\leftrightarrow$ neutral) charge state conversion processes, which are recognized to be stimulated by photo-excitation \cite{manson2005photo,siyushev2013optically}. 

In contrast to the findings in \cite{drake2015influence,loretz2017optical}, our experimental data, up to 83 mW/mm$^2$, does not exhibit any signs of saturation. This excludes a significant photo-ionization of (negatively charged) NV charge state in our sample. Conversely, our data reveals a higher experimental spin polarization for SQTs compared to that predicted by kinetic model calculations (even without considering photo-ionization processes) using typical values of different parameters reported in the literature, as discussed in figure S5 in the supplementary information.

Lastly, we analyze our experimental findings for both SQTs and DQT in the framework of seven-level kinetic model calculations. To start with, zero-field optical decay rates, including the intersystem crossing, for NV spin have been experimentally measured by different groups \cite{breeze2018continuous,manson2006nitrogen,robledo2011spin,tetienne2012magnetic,gupta2016efficient}. Specific values for different models are provided in Table S1 in supplementary information. A comprehensive analysis of the experimental results for different models are provided in figure S5 in the supplementary information. This reveals that the model 3, adapted from \cite{robledo2011spin}, describes well the angular ($\theta$) dependence for SQTs. 
% By varying the optical rates within a reasonable bound (10 \%) as provided in table \ref{tab:I}, we could find a good quantitative fit between our measured and computed spin polarizations for both SQTs and DQT.
However, when model 3 is used for DQT, although calculated values shows a similar trend to the experimental data, quantitatively, they tend to overestimate the spin polarization. This discrepancy can be attributed to the significantly lower $T_{1}$ value for DQT compared to SQTs, as previously reported by Sangtawesin et al. \cite{sangtawesin2019origins}.

In summary, our comprehensive study of NV transitions under off-axis fields has significant implications for advancing NV sensing for a broad variety of emerging quantum materials based on correlated electron systems as well as biological samples. To our knowledge this is the first experimental observation of the amplification of the double quantum transitions in NV centers and we likely have also seen the amplification of the signal from the nuclear double quantum transitions as well. These findings can also offer valuable insights into nuclear hyperpolarization using NV centers.

\section{Methods}\label{sec:methods}

The study employed a commercial continuous-wave Electron Spin Resonance (ESR) Spectrometer (JEOL JES-FA200) operating in the X-Band ($\sim$ 9.43 GHz) with a cylindrical TE\textsubscript{011} cavity. The (100)- single crystal diamond (DNV-B1 from Element Six, NV concentration: ~ 200 ppb) was fixed to a quartz tube at $45^\circ$ and placed in the cavity such that the $\braket{110}$ axis of the diamond crystal is parallel to the quartz tube. The quartz tube was then rotated about its axis (along the lab vertical) to change the orientation of the NV-axis w.r.t the applied static field. This ensures that 2 of the 4 possible orientations of NV in diamond are always perpendicular to the microwave field and the applied magnetic field stays in the [110] plane. The angle was adjusted manually using a goniometer with 1$^\circ$ resolution. The sample was illuminated by a 532 nm laser via an optical access to the cavity. All the ESR measurements were performed at room temperature and with a low microwave power of 10 $\mu$W, after verifying the linear response of the ESR signal at this power level.

\section*{ACKNOWLEDGEMENTS}

This work was supported by DST, India, under QuST program vide sanction no. DST/ICPS/QuST/Theme-2/2019/General and IIT Madras via exploratory research and team research grants. The authors extend their thanks to the staff at Sophisticated Analytical Instruments Facility (SAIF), IIT Madras for making the EPR facility available for our use. 

\bibliography{Ref_NV_ESR_Article}% Produces the bibliography via BibTeX.
\end{document}